\documentclass[aps,prl,twocolumn,superscriptaddress]{revtex4}

\usepackage{graphicx}

\newcommand{\figwidth}{3.375in}

\newcommand{\eq}[1]{Eq.~(\ref{#1})}

\begin{document}

\title{A real space renormalization group approach to spin glass
  dynamics}

\author{Falk Scheffler}
\affiliation{
Fachbereich Physik, Universit\"at Konstanz, 78457 Konstanz, Germany}
\author{Hajime Yoshino}
\affiliation{
Department of Earth and Space Science, Faculty of Science,
Osaka University, Toyonaka, 560-0043 Osaka, Japan}
\author{Philipp Maass}
\affiliation{
Institut f\"ur Physik,
Technische Universit\"at Ilmenau, 98684 Ilmenau, Germany}

\date{February 16, 2003}

\begin{abstract}
  The slow non--equilibrium dynamics of the Edwards--Anderson spin glass
  model on a hierarchical lattice is studied by means of a
  coarse--grained description based on renormalization concepts. We
  evaluate the isothermal aging properties and show how the occurrence
  of temperature chaos is connected to a gradual loss of memory when
  approaching the overlap length. This leads to rejuvenation effects
  in temperature shift protocols and to rejuvenation--memory effects in
  temperature cycling procedures with a pattern of behavior parallel
  to experimental observations.
\end{abstract}

\maketitle

Almost all glassy systems exhibit aging effects \cite{BCKM} which
reflect a slowing down of the dynamics. Spin glasses exhibit the
unusual feature that even small changes of temperature let them appear
as if they had experienced a much shorter thermalization period, which
is known as rejuvenation. Puzzlingly, they on the other hand keep
quite accurate memories of their past thermal histories
\cite{Saclay,Uppsala}. These phenomena call for a consistent
theoretical explanation.

In this Letter, we construct a simple yet powerful coarse--grained
approach for the relaxational dynamics of the Edwards--Anderson (EA)
Ising spin glass model on the hierarchical lattice associated with the
Migdal-Kadanoff (MK) approximation.  This approach allows us to
incorporate the exact real space renormalization group (RSRG)
transformations \cite{MKRG,ABM} into an effective non--equilibrium
dynamics on large time and length scales, which remain out of reach by
the conventional Monte Carlo technique.  By employing computer
simulations of the effective dynamics and analytical approaches
combined with scaling arguments we show how temperature chaos
\cite{BM,FH1,MKRG,ABM} progressively rejuvenates the system after
temperature shifts and how memory effects emerge after temperature
cycling.

\begin{figure}[b!]
\includegraphics[width=\figwidth]{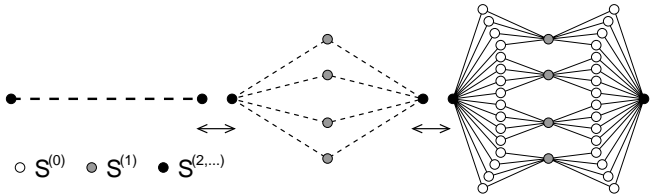}
\caption{Sketch of the hierarchical lattice. The iterative
construction corresponds to going from the left to the right in the
figure. Each bond in one iterative step is replaced by four pairs of
bonds with a new Ising spin in between.}
\label{fig:lattice}
\end{figure}

The hierarchical lattice of the MK approximation for the EA model in
three dimensions is constructed iteratively as depicted in
Fig.~\ref{fig:lattice}. To each bond in this lattice, a random
exchange interaction is assigned drawn from a Gaussian distribution
with zero mean and variance $J^2$.

The hierarchical construction of the lattice implies a hierarchy of
time scales in the system's relaxation. The spins $\{S^{(0)}\}$ with
only two neighbors introduced in the last iteration step belong to the
lowest level of the hierarchy and relax most quickly. Spins
$\{S^{(n)}\}$ at increasingly higher levels $n$ in the hierarchy have
an exponentially increasing number $2\times 4^{n}$ of nearest
neighbors and relax with correspondingly slower rates. In the MK
renormalization approach the spins $\{S^{(n)}\}$ are associated with
an exponentially increasing length scale $L_n=2^{n}L_0$ where $L_0$ is
the lattice constant.  For temperatures $T$ below the spin glass
transition temperature $T_c\simeq 0.88J$ \cite{ABM}, the effective
energy barrier $E_n$ for excitations (``flips'' of domains or droplets
\cite{FH1}, see below) of length scale $L_n$ will scale as
$E_n\sim J(L_n/L_0)^{\psi}$ with some exponent $\psi$.  Hence we
define the relaxation time $t_n$ for spins at level $n$ as
\begin{equation}
t_n/\tau_0=\exp\bigl[(J/T)(L_n/L_0)^{\psi}\bigr]=
\exp\bigl[2^{n\psi}(J/T)\bigr]\,,
\label{eq:t-epoch}
\end{equation}
where $\tau_0$ is a microscopic time unit. The time period $t_n$ is
denoted as the $n$th epoch following \cite{FH1}.

Our effective dynamics proceeds by successive epochs. Since these grow
with level $n$ exponentially even on a logarithmic time axis, we
consider, in the $n$th epoch, the spins $\{S^{(n+1,n+2,\ldots)}\}$ to
be frozen in and the spins $\{S^{(n,n-1,\ldots,0)}\}$ to fluctuate
with strongly decreasing relaxation times $t_n\gg t_{(n-1)}\gg
\dots\gg t_0$. In the $n$th epoch, we then first thermalize (align
with Boltzmann weights) the spins $S_i^{(n)}$ in their effective
(time--averaged) local fields $h_i^{(n)}\!=\!\sum_j
J_{ij}^{(n)}S_j^{(n+1,n+2,\ldots)}$. Here $J_{ij}^{(n)}$ are the
effective couplings given by the RSRG transformation \cite{MKRG,ABM} $
J^{(n)}_{ij}\!=\!T \sum_{k=1}^{4}\!\tanh^{-1} [
\tanh(J^{(n-1)}_{ik}/T) \tanh(J^{(n-1)}_{jk}/T) ]\,, $ which take into
account the thermalization of the faster spins at lower levels $n'<n$.
In the second step, the spins $S_i^{(n-1)}$ are thermalized in their
effective local fields $h_i^{(n-1)}$, which depend on the spins
$S_i^{(n)}$ updated in the first step (and the effective couplings
$\{J^{(n-1)}\}$).  By repeating this procedure, the spins
$\{S^{(n)}\}, \{S^{(n-1)}\}, \dots, \{S^{(0)}\}$ are updated one after
the other in the $n$th epoch.

To characterize the non--equilibrium properties of the system we consider a
quench from $T=\infty$ to $T<T_c$ at time zero and study the spin
autocorrelation function
\begin{equation}
C(t_m,t_n)=\sum_{\alpha} w_{\alpha} \sum_{i_{\alpha}}
\overline{\left\langle
S_{i_\alpha}^{(\alpha)}(t_n+t_m)\, 
S_{i_\alpha}^{(\alpha)}(t_n)\right\rangle}\,,
\label{eq-c-def}
\end{equation}
where $\langle\ldots\rangle$ denotes a thermal average and an average
over random initial spin orientations for a fixed realization of the
disorder, while the bar denotes the disorder average over the random
bonds. The $w_{\alpha}>0$, $\sum_\alpha w_\alpha=1$, are weighting
factors, which allow one to take into account that spins at different
levels are not equivalent. A natural choice is to take $w_\alpha$
proportional to the connectivity $\propto 4^{\alpha}$ of spins
$\{S^{(\alpha)}\}$ and we present our results for this case. We have
checked, however, that the choice $w_{\alpha}=const.$ \cite{MC} yields
analogous results. Recently, direct experimental measurements of the
correlation function have been conducted in a spin glass \cite{HO}.

The time evolution of the system is best analyzed in terms of
clusters, which are realizations of domains or droplets postulated in
scaling arguments \cite{FH1}. A cluster is defined for each effective
coupling $J^{(\alpha)}$ and consists of all faster spins at levels
below $\alpha$, which are traced out in the RSRG to give
$J^{(\alpha)}$. Each cluster has two boundary spins which are the
spins connected by $J^{(\alpha)}$ after renormalization. One is a spin
$S^{(\alpha)}$ at level $\alpha$, which we call the ``master spin'' of
the cluster. The other is a still slower spin of level
$\gamma>\alpha$.  An important remark is that $2J^{(\alpha)}$ is the
difference of the cluster's free energies of the cases that the
boundary spins are parallel and anti--parallel. If the master spin
flips but the slower boundary spin keeps unchanged, the spins in the
interior of the cluster are exposed to a ``twisted--boundary
condition'', which triggers flips of a $O(1)$ fraction of these spins.
It is by this mechanism that fluctuations occurring at high levels
propagate down and erase correlations of the low--level spins.

In fact, these de--correlations caused by twisted boundaries can be
quantified in a precise manner. Let us consider a certain spin $\hat
S^{(\alpha)}$ at level $\alpha$. Since the clusters are hierarchically
nested, this spin $\hat S^{(\alpha)}$ is part of unique clusters with
master spins $\hat S^{(\gamma)}$, $\gamma>\alpha$. By using symmetry
considerations we can write for the correlator appearing in
\eq{eq-c-def}
\begin{eqnarray}
\label{c-cluster-r-rep}
\overline{\left\langle
\hat S^{(\alpha)}(t_n+t_m)\, 
\hat S^{(\alpha)}(t_n)\right\rangle}
&=&\\
&&{}\hspace*{-10em}
\overline{[1-2r_\alpha(t_m,t_n)]
\prod_{\gamma=\alpha+1}^{m}[1-r_\gamma(t_m,t_n)]}\,,
\nonumber
\end{eqnarray}
where $r_\gamma(t_m,t_n)$ is the probability that the spin $\hat
S^{(\gamma)}$ flips between $t_n$ and $t_n+t_m$ under the condition
that the master spins $\hat S^{(\gamma)}$ remained unchanged (for a
given realization of the disorder). The product over $\gamma$ in
\eq{c-cluster-r-rep} is the consequence of the twisted boundary
effect: None of the boundary spins of the ``super''--clusters
containing the spin $\hat S^{(\alpha)}$ is allowed to flip between
$t_n$ and $t_n+t_m$, if $S^{(\alpha)}(t_n+t_m)$ should give a non--zero
overlap with $S^{(\alpha)}(t_n)$.  For small flip probabilities
$r_\gamma\ll1$, the right hand side of \eq{c-cluster-r-rep} can
be linearized, yielding $1-2\bar
r_\alpha-\sum_{\gamma=\alpha+1}^{m}\bar r_\gamma$.

\begin{figure}[t!]
\includegraphics[width=3.0in]{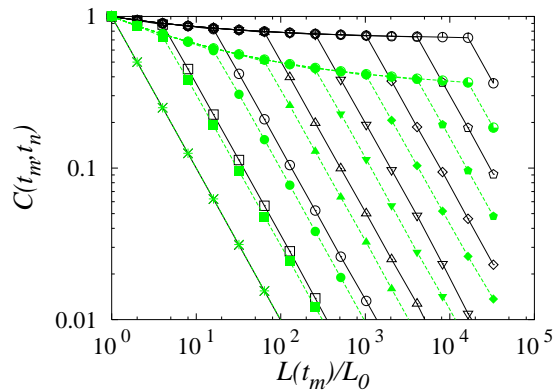}
\caption{Spin autocorrelation function $C(t_m,t_n)$ as a function
  of $L(t_m)$ at $T/T_{c}=0.3$ (open symbols) and $T_{c}=0.7$
  (filled symbols) for $L(t_n)/L_0=2^{0},2^{2},\ldots,2^{14}$ (from
  left to right). Here and in the following, the system size was
  $2^{15}L_0$.}
\label{fig:isothermal-shift}
\vspace*{-3ex}
\end{figure}

{\it Isothermal aging --} Let us first discuss the isothermal aging
properties. Fig.~\ref{fig:isothermal-shift} shows the simulated
$C(t_m,t_n)$ as a function of $L(t_m)/L_0\!  =\![(T/J)
\log(t_m/\tau_0)]^{1/\psi}$ [cf.\ \eq{eq:t-epoch}].  Clearly,
there are two different regimes, a quasi--equilibrium regime with a
slow decay for $L(t_m)\le L(t_n)$ and the aging regime with a fast
decay for $L(t_m)>L(t_n)$.

Relaxation in the quasi--equilibrium regime $L(t_m)\le L(t_n)$ is close
to that of the equilibrium limit $C_{\rm eq}(t_m)
=\lim_{n\to\infty}C(t_m,t_n)$. In the low--$T$ limit the behavior of
$C_{\rm eq}(t_m)$ can be derived by considering thermal fluctuations
from the ground states. At $T=0$ a spin $S_i^{(\alpha)}$ points in the
direction of the effective local field $h_i^{(\alpha)}$.  Accordingly,
for $T\gtrsim 0$ spin flips become likely if the energy gap
$\Delta_\alpha=2|h_i^{(\alpha)}|$ is smaller than the thermal energy
$T$.  We have investigated numerically the distribution
$\rho(\Delta_\alpha)$ of the energy gaps and found that it follows an
analogous scaling form as the distribution of renormalized bonds
\cite{FH1}
\begin{equation}
\rho(\Delta_\alpha) d \Delta_\alpha = 
\tilde{\rho}\left(\frac{\Delta_\alpha}{J(L_\alpha/L_0)^{\theta}}\right) 
\frac{d \Delta_\alpha}{J(L_\alpha/L_0)^{\theta}}\,, 
\label{eq-scaling-gap}
\end{equation}
where $\tilde{\rho}(0)>0$ and $\theta\simeq 0.26$ \cite{ABM} is the
stiffness exponent. As a consequence, we obtain up to terms of order
$O(T^2,L_\alpha^{-1-\theta})$ for the disorder averaged flip
probabilities $\bar r_\gamma\sim c(T/J)(L_\gamma/L_0)^{-\theta}\ll1$
with $c=\tilde\rho(0)/2$. Using the linearized form of
\eq{c-cluster-r-rep} this yields
\begin{equation}
C_{\rm eq}(t_m)\sim 
q_{\rm EA} + c'\,\frac{T}{J}\left(\frac{L(t_m)}{L_0}\right)^{-\theta},
\label{eq:c-isotherm}
\end{equation}
where $c'=c/(2^\theta-1)$ and the Edwards--Anderson order parameter
$q_{\rm EA}$ decreases linearly with $T$, $q_{\rm EA}\sim
1-const.\,T$. Indeed, we could well fit the data in
Fig.~\ref{fig:isothermal-shift} to \eq{eq:c-isotherm} and thereby
extract $q_{\rm EA}$, which we found to decrease linearly with $T$
except possibly close to $T_c$.

In the aging regime $L(t_m)>L(t_n)$, the master spins at levels
$n<\gamma\le m$ are newly thermalized and flip with probability
$r_\gamma=1/2$. Taking this into account in
\eq{c-cluster-r-rep} we find the exact relation
\begin{equation}
C(t_m, t_n)=
C(t_n,t_n)\left(\frac{L(t_m)}{L(t_n)}\right)^{-\lambda}
 \label{eq-iso-age}
\end{equation} 
with $\lambda=1$. Equations~(\ref{eq:c-isotherm},\ref{eq-iso-age}) agree with
the scaling forms suggested by the droplet scaling theory \cite{FH1}.

\begin{figure}[b!]
\includegraphics[width=3.0in]{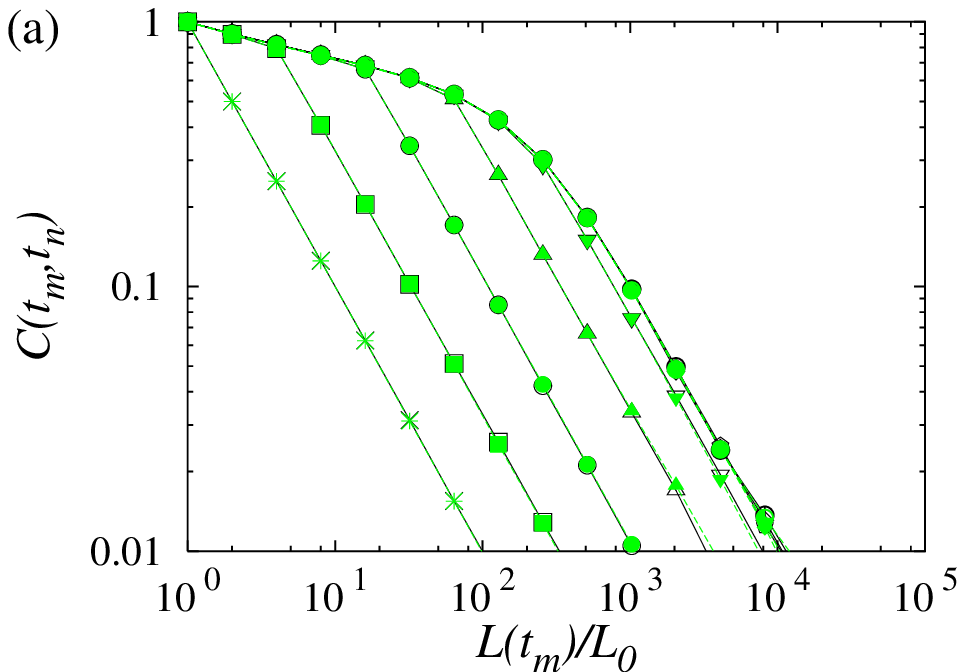}
\includegraphics[width=3.0in]{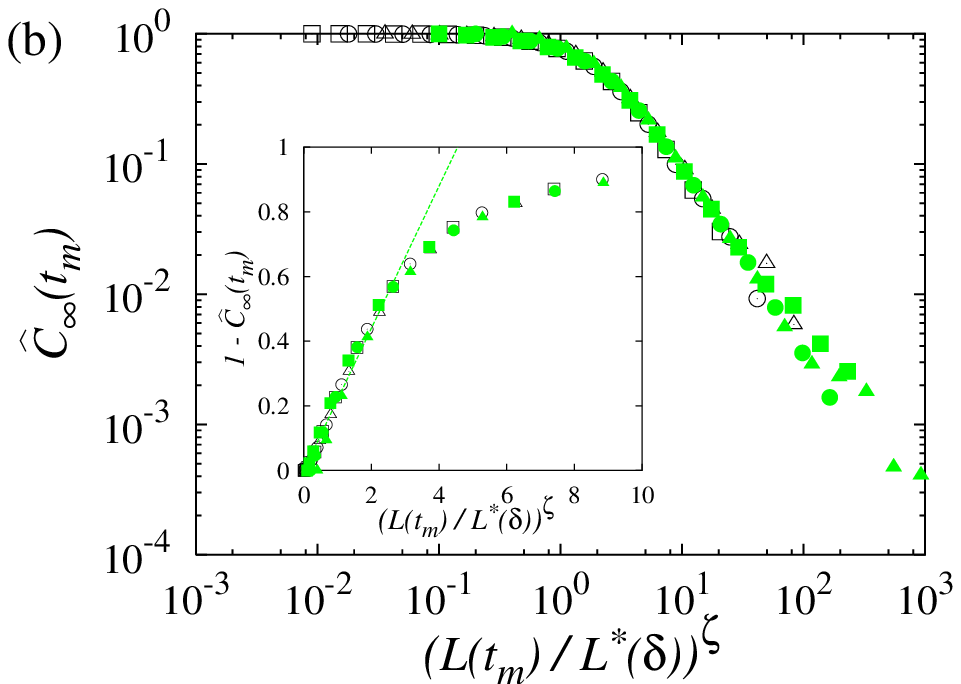}
\caption{{\it (a)} Spin autocorrelation function $C(t_m,t_n)$ as a function
  of $L(t_m)$ in twin temperature shift protocols $(0.3 T_{c},0.7
  T_{c})$ (open symbols) and $(0.7 T_{c},0.3 T_{c})$ (filled symbols)
  for $L(t_n)/L_0=2^{0},2^{2},\ldots,2^{14}$ (from left to right).
  According to \cite{ABM}, the overlap length between the two
  temperatures is $L^*/L_0=2^{7}-2^{8}$.  {\it (b)} Scaling plot of
  the normalized autocorrelation function $\hat C_\infty(t_m;\delta)$
  as a function of $[L(t_m)/L^*(\delta)]^\zeta$ with $\zeta=0.74$
  \cite{ABM}. The open/filled symbols are for temperature/bond shifts
  with $\delta=0.1$ (square), $0.2$ (circle) and $0.5$ (triangle).
  The $T$--shifts started from $T_1=0.3T_c$, while the bond shifts were
  in the $T=0$--limit of the dynamics. The prefactor of $L^*(\delta)$
  is chosen such that the two master curves for temperature and bond
  shifts merge with each other. The inset shows the initial decay of
  $\hat C_\infty(t_m;\delta)$ in agreement with \eq{eq:shift-c-st}.}
\label{fig:T-shift}
\end{figure}

{\it Temperature/bond shift protocols --} Next we proceed to our
central issue and investigate the implication of temperature chaos on
aging and rejuvenation effects in temperature shift protocols
$(T_1,T_2)$. The system evolves first at temperature $T_1$ for time
$t_n$ and then the temperature is switched to $T_2=T_1+\Delta T$,
corresponding to a perturbation of strength $\delta\equiv|\Delta
T|/T_c$.

Figure~\ref{fig:T-shift}a shows the correlation function $C(t_m,t_n)$
as a function of $L(t_m)$ obtained from our computer simulations. As
in Fig.~\ref{fig:isothermal-shift}, we can identify an aging regime
for $L(t_m)>L(t_n)$. Indeed the scaling from \eq{eq-iso-age} holds
exactly with the same underlying mechanism as in the isothermal case.
In this respect the aging effect does not stop. However, the limiting
curve $C_{\infty}(t_m)=\lim_{n\to\infty}C(t_m,t_n)$ decays to zero for
$t_m\to\infty$ while that of isothermal aging converges to the plateau
$q_{\rm EA}$ [see \eq{eq:c-isotherm}]. Hence the aging effect becomes
progressively irrelevant and the system rejuvenates, suggesting that
temperature chaos effect is coming into play.

Another remarkable observation from Fig.~\ref{fig:T-shift}a is that
$C(t_m,t_n)$ remains unchanged when $T_1$ and $T_2$ are interchanged.
Due to symmetry this must hold true for $t_m=t_n$, and, because of
(\ref{eq-iso-age}), it is exactly valid for $t_m>t_n$ also. For
$t_m<t_n$, it can be shown based on \eq{c-cluster-r-rep} that
deviations, if occurring at all, must be very small.  The symmetry
supports the results of recent ``twin--experiments'' performed on a
spin glass \cite{JYN2002}.

To understand the behavior of $C_{\infty}(t_m)$ it is important to
note that a temperature shift affects the spin flip probabilities in
two ways: {\it (i)} it changes the effective couplings due to the
$T$--dependence of the RSRG transformations and hence the effective
fields, and {\it (ii)} it changes the weighting of these fields in the
Boltzmann probabilities.  In order to isolate the subtle effect {\it
  (i)} from the obvious effect {\it (ii)} we introduce a correlation
function that is normalized by the isothermal correlation functions at
temperatures $T_1$ and $T_2$, $\hat C_\infty(t_m) \equiv
C_\infty(t_m)/\sqrt{C_{\rm eq}(t_m)_{T_1} C_{\rm eq}(t_m)_{T_2}}$. It
can be shown that $\hat C_\infty=1+O(\delta^2)$ if effect {\it (i)} is
absent.

Now we can focus on effect {\it (i)} and estimate the induced flip
probability $\bar r_\alpha(\delta)$ of a master spin $\hat
S^{(\alpha)}$.  A flip of such a spin becomes likely if the typical
change of effective couplings $\Delta J^{(\alpha)}$ induced by the
temperature shift exceeds the original energy gap $\Delta_\alpha$
before the shift, i.e.\ $\bar r_\alpha(\delta)\sim \int_0^{\Delta
  J^{(\alpha)}}d \Delta_\alpha \rho(\Delta_\alpha)$. As noted above,
$2J^{(\alpha)}$ is the difference of the free energies of the
associated cluster between the two states with parallel and
anti--parallel boundary spins, which below $T_{c}$ have a ``relative
domain wall'' that passes through $L_{\alpha}^{d-1}$ links.
Accordingly, one can write for the free energy change $2\Delta
J^{(\alpha)}\sim \Delta[TS_{\alpha}]$, where $S_{\alpha}$ is the
difference of the entropies associated with the two states.  As shown
in \cite{FH1,BM,ABM}, $S_{\alpha}$ is the sum of random local entropy
fluctuations associated with the $L_\alpha^{d-1}$ links along the
relative domain wall, yielding $S_{\alpha}\sim L_\alpha^{(d-1)/2}$.
It follows that $\Delta J^{(\alpha)}\sim \delta L_\alpha^{(d-1)/2}$
and hence $\bar r_\gamma(\delta)\sim
\int_0^{[L_\gamma/L^*(\delta)]^{\zeta}} dy \tilde{\rho}(y)$, where
$L^*(\delta)/L_0 \sim \delta^{-1/\zeta}$ is the so--called overlap
length. The exponent $\zeta=(d-1)/2-\theta\simeq 0.74$ is called the
chaos exponent \cite{BM,FH1}.

In the {\it weakly perturbed regime} $L(t_{m})\ll L^{*}(\delta)$ the
flip rates are small, $\bar r_\gamma(\delta)\sim \tilde{\rho}(0)
[L_\gamma/L^*(\delta)]^{\zeta}$, and using the linearized
\eq{c-cluster-r-rep} we obtain
\begin{equation}
{\hat C}_{\infty}(t_m;\delta)=1 - c \tilde{\rho}(0)\left(
\frac{L(t_{m})}{L^{*}(\delta)} \right)^{\zeta}
\label{eq:shift-c-st}
\end{equation}
where $c$ is a constant. Quite importantly this means that the
rejuvenation effect does not appear suddenly at the overlap length
$L^{*}(\delta)$ but rather gradually emerges. Very similar scaling for
the emergence of the rejuvenation effect is found in the
``twin--experiment'' mentioned above \cite{JYN2002}.

In the late stage $L(t_{m}) \gg L^{*}(\delta)$, which we call {\it
  strongly perturbed regime}, the flip rates become $1/2$. Then
similarly to \eq{eq-iso-age} we expect,
\begin{equation}
{\hat C}_{\infty}(t_{m};\delta)
\sim \left(\frac{L(t_{m})}{L^{*}(\delta)} \right)^{-\lambda} 
\label{eq:shift-c-lt}
\end{equation}
with $\lambda=1$. 

Equivalent effects are expected for bond shifts \cite{FH1,BM}.  In a
bond shift, the bare couplings $J_{ij}$ are replaced by
$(J_{ij}+\delta J'_{ij})/\sqrt{1+\delta^{2}}$, where $J'_{ij}$ are
Gaussian random numbers obeying the same statistics as the $J_{ij}$,
and $\delta$ is the perturbation strength. 

We tested the above scaling ansatz numerically. As demonstrated in
Fig.~\ref{fig:T-shift}, they work very well. Quite remarkably the form
of the master curve is apparently universal for both temperature
and bond shifts.

\begin{figure}[t!]
\includegraphics[width=3.0in]{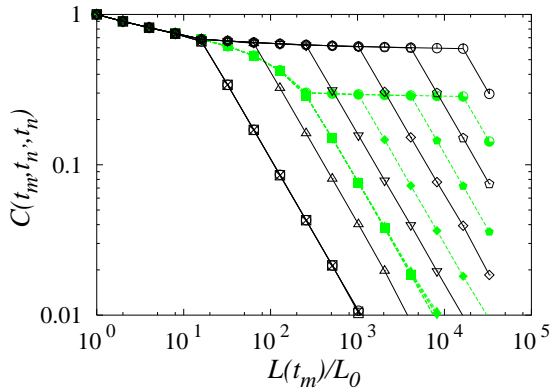}
\caption{Rejuvenation and memory: relaxation of the autocorrelation
  function after one--step temperature cycling $T_1 \to T_2 \to T_1$
  with $T_1=0.3T_c$ and $T_2=0.7T_c$.  The duration of the first stage
  is varied as $L(t_{n})/L_0=2^{0},2^{2},\ldots,2^{14}$ from left to
  right.  The duration of the second stage is $L(t_{n'})/L_0=2^{4}$ (open
  symbols) and $2^{8}$ (filled symbols).}
\label{fig:t-cycling}
\vspace*{-3ex}
\end{figure}

{\it Temperature cycling--} Finally, we study rejuvenation and memory
effects after one step temperature cycling. Therein, the system
evolves at a temperature $T_1$ for a time $t_{n}$ and subsequently at
another temperature $T_2$ for a time $t_{n'}$. Then the temperature is
put back to $T_1$ and we measure the autocorrelation function
$C(t_{m},t_{n'},t_{n})$, which is the overlap between the spin
configuration at time $t_{n}+t_{n'}$ and that after some additional
time $t_{m}$.

Figure~\ref{fig:t-cycling} shows the simulated data. The initial decay
shows rejuvenation: There is no trace of aging during the first stage
and $C(t_{m},t_{n'},t_{n})$ follows the autocorrelation function
$C(t_{m}, t_{n'}; \delta= |T_1-T_2|/T_{c})$ as after a simple
$T$--shift $(T_2,T_1)$. If $L_{T_1}(t_{n})>L_{T_2}(t_{n'})$, this
rejuvenated, new aging is however terminated after the recovery time
$t_{m^\star}$ determined by $L_{T_1}(t_{m^\star}) =L_{T_2}(t_{n'})$
\cite{YLB}. There, a plateau region shows up indicating that the
system's evolution recovers length scales already thermalized at
temperature $T_1$ during the first stage. Eventually, for
$t_{m}>t_{n}$, the system also memorizes the limits of this first
thermalization and $C(t_{m},t_{n'},t_{n})$ enters a steep decay
analogous to \eq{eq-iso-age} in the isothermal case. We confirmed that
bond cycling yields analogous results.

The two stage relaxation is reminiscent of the experimentally observed
rejuvenation--memory effects \cite{Saclay,Uppsala}.  Here slow spins
play the role of ghost domains proposed in \cite{YLB}.  After strong
perturbations, configurations of faster spins which occupy a dominant
portion of the volume of a cluster are changed  but slower
spins in the same volume retain their original configuration. The
slower spins act as remanent symmetry breaking fields by which both
the original {\it amplitude} and {\it sign} of the overlap with
respect to the equilibrium state can be restored.

F.S.\ thanks the German Academic Exchange Service (DAAD) and the
Japanese Ministry of Science (MEXT) for financial support during a
stay in Japan. H.Y.\ is supported by the Ministry of Education,
Culture, Sports, Science and Technology of Japan, Grant--in--Aid for
Scientific Research 14740233.

\end{document}